\documentclass[a4paper]{article}

\usepackage{INTERSPEECH2020}
\usepackage{subfig}
\usepackage{cite}
\usepackage{multirow}

\title{Dual Attention in Time and Frequency Domain for Voice Activity Detection}
\name{Joohyung Lee, Youngmoon Jung, Hoirin Kim}
\address{
  School of Electrical Engineering, KAIST, Daejeon, South Korea}
\email{\{wngud701,dudans,hoirkim\}@kaist.ac.kr}

\begin{document}

\maketitle
\begin{abstract}
Voice activity detection (VAD) is a challenging task in low signal-to-noise ratio (SNR) environment, especially in non-stationary noise. To deal with this issue, we propose a novel attention module that can be integrated in Long Short-Term Memory (LSTM). Our proposed attention module refines each LSTM layer's hidden states so as to make it possible to adaptively focus on both time and frequency domain. Experiments are conducted on various noisy conditions using Aurora 4 database. Our proposed method obtains the 95.58 \% area under the ROC curve (AUC), achieving 22.05 \% relative improvement compared to baseline, with only 2.44 \% increase in the number of parameters. Besides, we utilize focal loss for alleviating the performance degradation caused by imbalance between speech and non-speech sections in training sets. The results show that the focal loss can improve the performance in various imbalance situations compared to the cross entropy loss, a commonly used loss function in VAD.
\end{abstract}

\section{Introduction}

Voice activity detection (VAD) is a kind of binary classification which classifies a frame into speech or non-speech. VAD is an important pre-processing step in speech applications such as automatic speech recognition (ASR), speaker recognition, speech enhancement, and speech coding, etc. The early approaches to VAD were based on signal processing-based approaches using time-domain power \cite{Rabiner1975}, zero crossing rate (ZCR) \cite{Junqua1992}, cepstral features \cite{Haigh1993}, or spectral entropy \cite{Shen1998}. Besides, statistical model-based approaches have been widely adopted using Gaussian models for speech and noise signals \cite{Sohn1999, Chang2006}.

Recently as deep learning has been on the rise, it has shown its effectiveness on finding the optimal VAD models such as fully-connected deep neural networks (FCDNNs) \cite{Zhang2014, Jung2017, Jung2018}, convolutional neural networks (CNNs) \cite{Thomas2014, Sehgal2018, Lin2019}, Long Short-Term Memories (LSTMs) \cite{Eyben2013, Kim2016, Sertsi2017}, and the combination of deep neural networks \cite{Zazo2016, Vafeiadis2019}. However, although those deep learning-based VAD models have achieved marked improvements, VAD is still a challenging task in low signal-to-noise ratio (SNR) environments. 

To improve the robustness in noisy environments, we propose a novel VAD model based on attention method. Our architecture is motivated by the attention module integrated to CNN architecture used in computer vision \cite{Hu2017, Woo2018}. These attention modules squeeze the intermediate feature map and increase representation power of networks. They substantially improve the performance with small overhead. Motivated by them, several studies have been conducted in speech signal processing \cite{Yadav2020, Yanpei2020}. However, to our best knowledge, there has been no attempt to apply them for VAD. Our proposed attention module squeezes the feature map by statistical pooling and adaptively focuses on important speech frames and frequency components in time and frequency domain, respectively.

Meanwhile, in supervised learning-based classification problem, class imbalance of training data can bring about deterioration since easily classified samples dominate the training procedure \cite{Japkowicz2002, He2009}. In case of VAD as well, audio samples in database usually show the imbalance between speech and non-speech sections. Indeed, cross entropy loss, broadly used in VAD, is not suitable for handling the class imbalance. On the other hand, a focal loss proposed in \cite{Lin2017} has a modulating term which is able to focus learning on minor samples in class imbalance situations. In the experiment, we utilize the focal loss in various class imbalance situations and demonstrate that it is conducive to class imbalance situations for VAD.

The remaining part of paper is organized as follows. Section 2 describes 4 types of proposed attention modules. Section 3 indicates problems about class imbalance in VAD and compares focal loss with cross entropy loss. Section 4 describes the experimental setup and Section 5 shows the results and analysis of experiments. Then, we conclude this work in Section 6.

\section{Attention Module}
Unlike original attention modules \cite{Hu2017, Woo2018}, we change the backbone architecture from CNN to LSTM. The reason of using LSTM instead of CNN is that LSTM shows the best performance among FCDNN, CNN, and LSTM in VAD with a similar number of parameters \cite{Tong2016, Wang2019}.
\begin{figure}[t!]
\centering{\includegraphics[clip, trim = {0 0 1cm 0.8cm}, width=\linewidth]{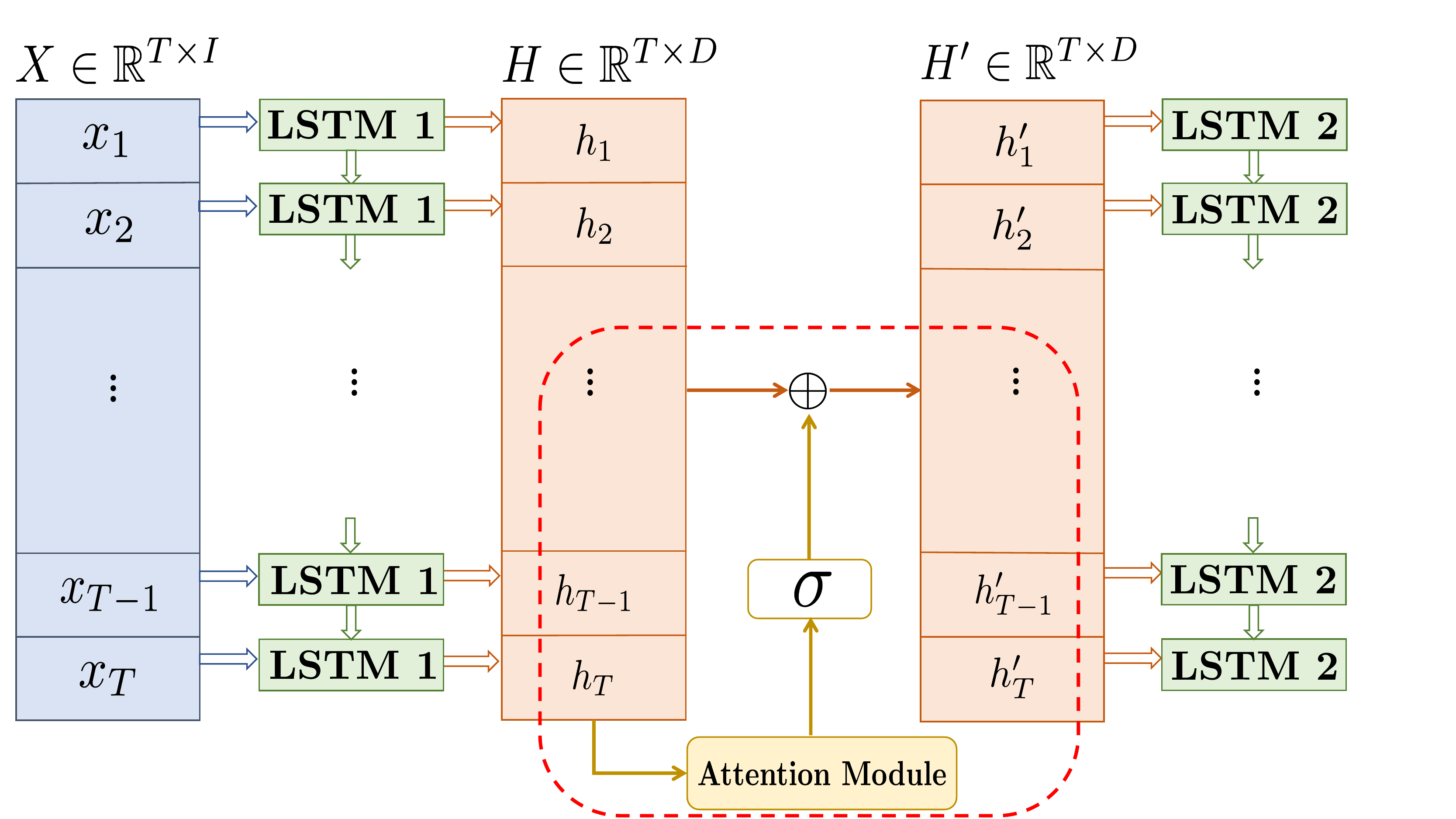}}
\vspace*{-0.6cm}
\caption{Illustration of proposed attention-based LSTM model.}
\label{fig:Proposed Model} 
\vspace{-0.5cm}
\end{figure}

\begin{figure*}
    \subfloat[\label{fig:TA}]{
        \includegraphics[clip, trim = {0 0 0 1.55cm}, width=0.285\linewidth]{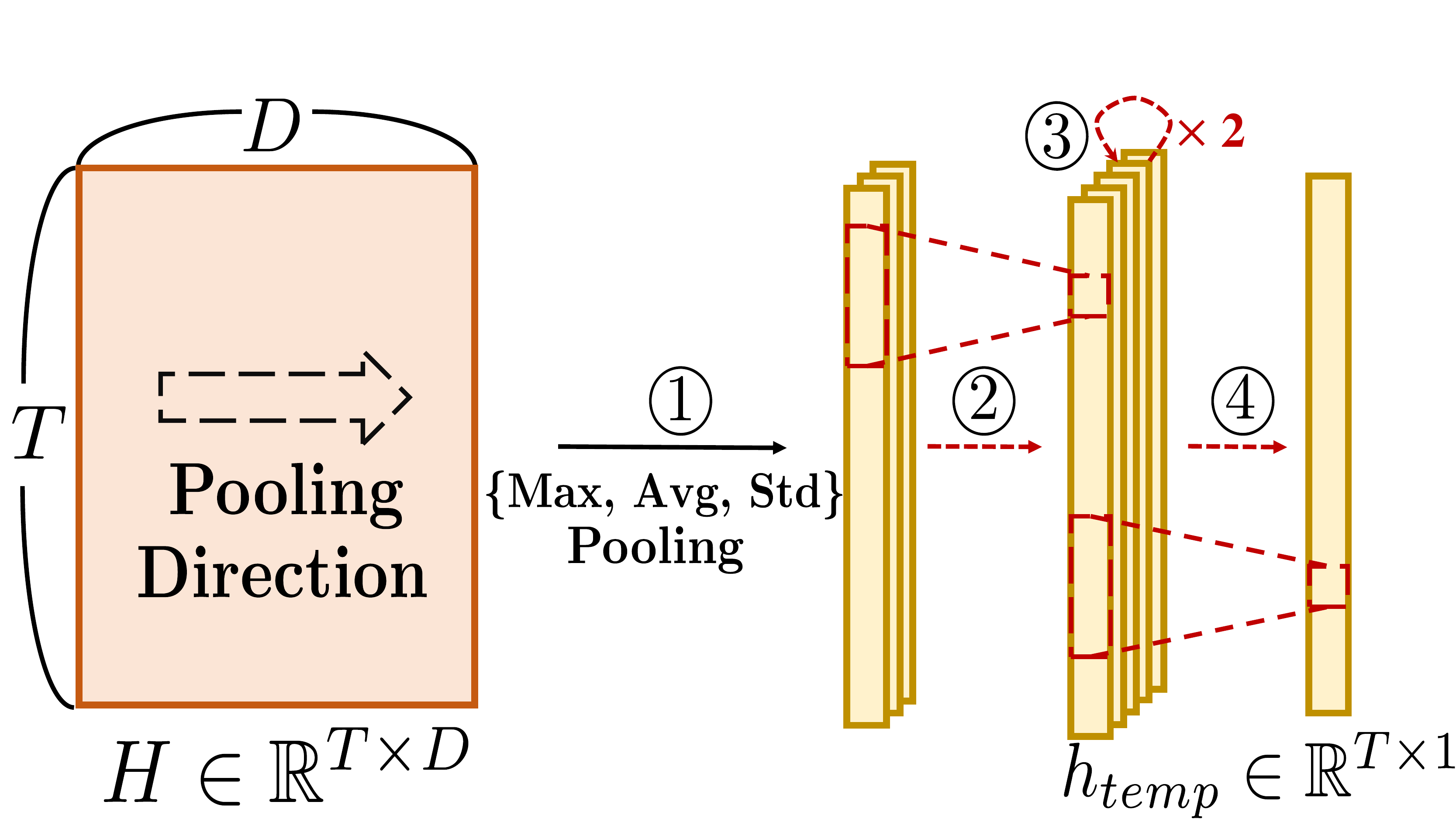}}
\hspace{\fill}
    \subfloat[\label{fig:FA}]{
        \includegraphics[clip, trim = {0 0 0 1.55cm}, width=0.285\linewidth]{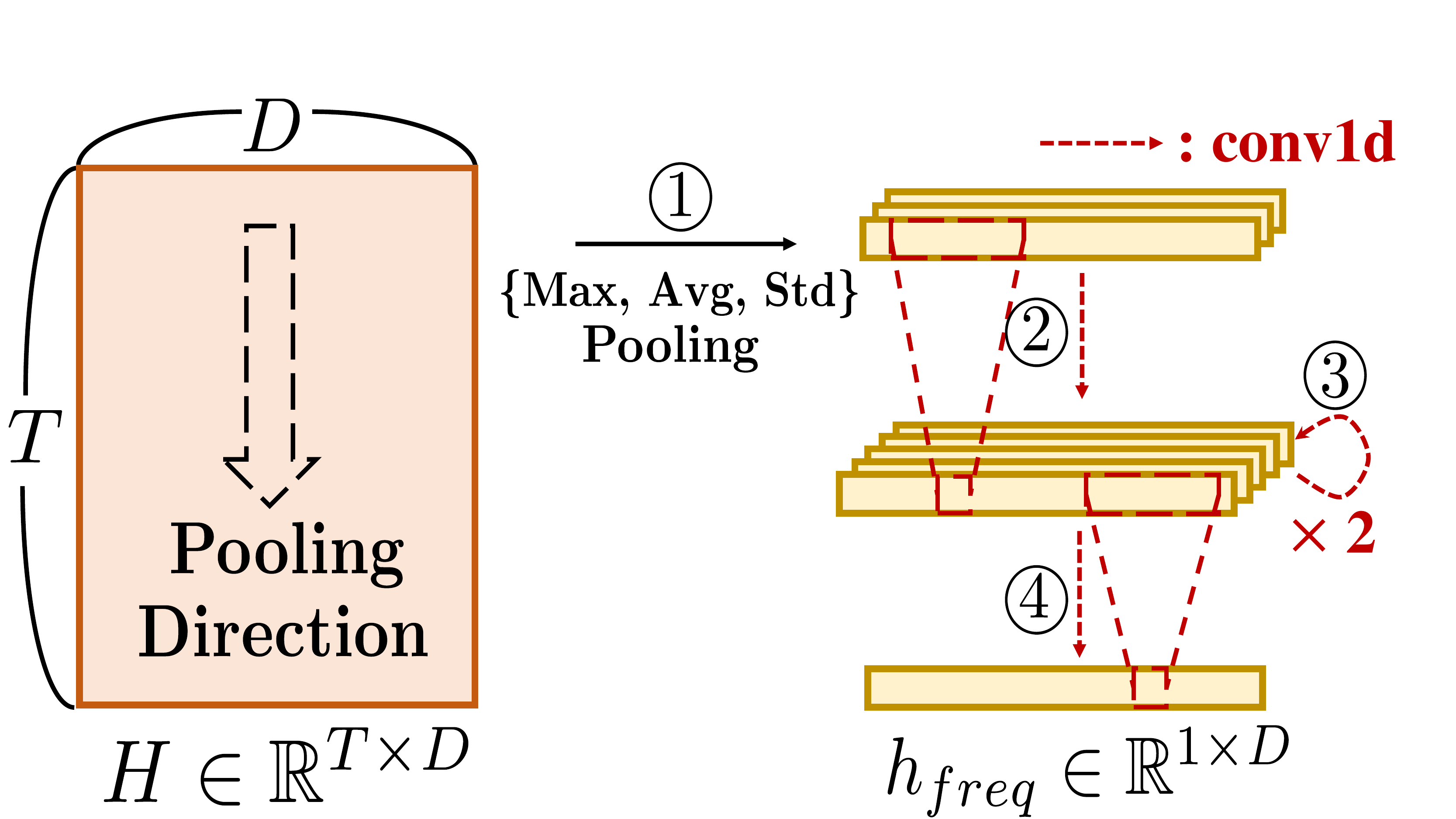}}
\hspace{\fill}
    \subfloat[\label{fig:DA-1}]{
        \includegraphics[clip, trim = {0 0 0 1.55cm}, width=0.285\linewidth]{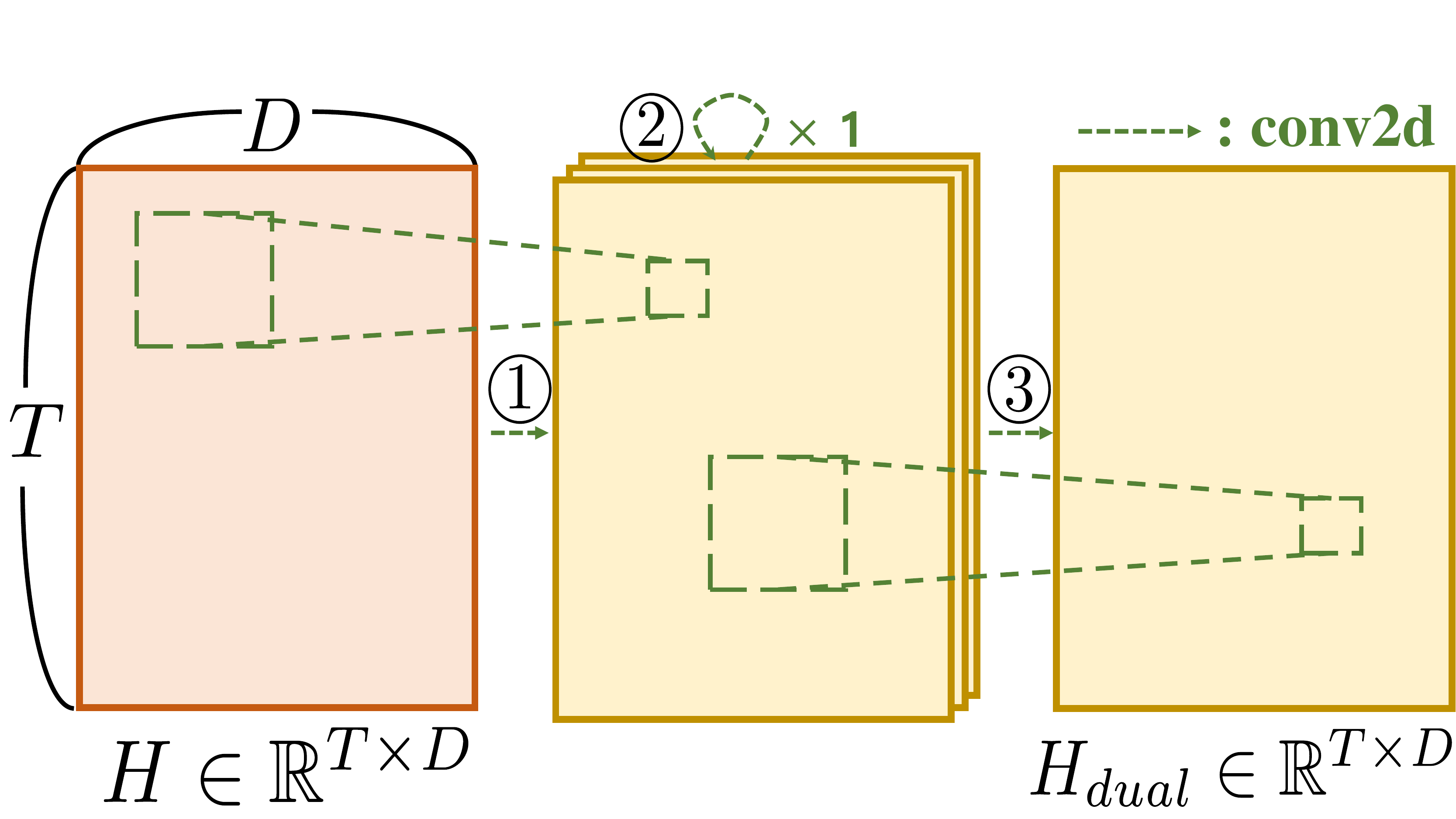}}
\vspace*{-0.3cm}
\caption{Illustration of proposed attention modules. (a) Temporal attention (TA); (b) Frequential attention (FA); (c) Dual attention 1 (DA-1); The circled number \textcircled{n} denotes the process of each attention module. $T$ denotes the sequence length and $D$ denotes the number of hidden nodes. $h_{temp}$ and $h_{freq}$, the output of temporal attention (Figure \ref{fig:TA}) and frequential attention (Figure \ref{fig:FA}), are expanded (copied) as $H$ before being activated by sigmoid function.}
\label{fig:Attention_Module}
\vspace{-0.5cm}
\end{figure*}

The structure of proposed attention-based LSTM model is shown in Figure \ref{fig:Proposed Model}. $X\in \mathbb{R}^{T \times I}$ is input acoustic features for model. $T$ denotes the length of time steps (sequence length) and $I$ denotes the dimension of acoustic features. When $X$ is fed into the first LSTM layer (LSTM 1), then hidden states $H \in \mathbb{R}^{T \times D}$ with $D$ hidden nodes are drawn. In basic LSTM, hidden states $H$ are fed to the next LSTM layer (LSTM 2) directly \cite{Sepp1997}. But in this paper, hidden states are refined as $H' \in \mathbb{R}^{T \times D}$ by proposed attention module before being fed to the next LSTM layer (LSTM 2). For refining hidden states, we regard hidden states $H$ as a kind of 2-dimensional \emph{hidden feature map}. In following subsections, we propose 4 kinds of attention modules.

\subsection{Temporal Attention (TA)}
\label{ssec_TA}

The temporal attention (TA) exploits temporal information and concentrates on important time steps for improving model's ability to discriminate speech frames from non-speech frames. Figure \ref{fig:TA} illustrates the process to obtain $h_{temp}$, output of TA module. The hidden feature map $H$ is pooled in three ways: max, average, and standard deviation pooling along the frequency axis, resulting $h_{temp}^{max}, h_{temp}^{avg}, h_{temp}^{std} \in \mathbb{R}^{T \times 1}$, respectively (step \raisebox{.5pt}{\textcircled{\raisebox{-.9pt} {1}}}). These 3 pooled feature vectors are concatenated and convolved by 1-dimensional convolution layers (step \raisebox{.5pt}{\textcircled{\raisebox{-.9pt} {2}}} - \raisebox{.5pt}{\textcircled{\raisebox{-.9pt} {4}}}). $h_{temp}$, the output of last convolution layer, is expanded (copied) as original hidden feature map $H$ (denoted as $H_{temp}$) and activated by sigmoid function. Finally, it is merged to $H$ by element-wise summation, then refined hidden feature map $H'$ is obtained. TA can be represented as below.
\begin{gather}
h_{temp} = f_{temp}([h_{temp}^{max};h_{temp}^{avg};h_{temp}^{std}])\label{eq:TA}
\\H' = H \oplus \sigma(H_{temp})\quad for\quad H_{temp} \in \mathbb{R}^{T \times D}\,,
\end{gather}
where $f_{temp}$ denotes the 1-dimensional convolution with 11 of kernel size in TA module. It consists of 4 layers and the number of filters is 3, 5, 5 and 1, respectively. $\sigma$ denotes the sigmoid function and $\oplus$ denotes the element-wise summation.

\subsection{Frequential Attention (FA)}
\label{ssec_FA}

The frequential attention (FA) is same with TA but for pooling direction and kernel size of convolution layer. Figure \ref{fig:FA} illustrates the process to obtain $h_{freq}$, output of FA module. $h_{freq}^{max}, h_{freq}^{avg}, h_{freq}^{std} \in \mathbb{R}^{1 \times D}$ are the max, average, and standard deviation pooling results of hidden feature map $H$ along the time axis (step \raisebox{.5pt}{\textcircled{\raisebox{-.9pt} {1}}}). Like in TA, these 3 feature vectors are concatenated, and then passed to convolution layers (step \raisebox{.5pt}{\textcircled{\raisebox{-.9pt} {2}}} - \raisebox{.5pt}{\textcircled{\raisebox{-.9pt} {4}}}) and sigmoid function. Also, after being expanded as $H$ (denoted as $H_{freq}$), it is merged to $H$ by element-wise summation for obtaining refined hidden feature map $H'$. FA can be represented as below.
\begin{gather}
h_{freq} = f_{freq}([h_{freq}^{max};h_{freq}^{avg};h_{freq}^{std}])\label{eq:FA}
\\H' = H \oplus \sigma(H_{freq})\quad for\quad H_{freq} \in \mathbb{R}^{T \times D} \,,
\end{gather}
where $f_{freq}$ denotes the 1-dimensional convolution with 21 of kernel size in FA module. Sequence length $T$ in training steps is fixed by pre-defined value to build the mini-batch for training. But in testing steps, the value of $T$ changes according to the length of audio sample. This mismatch of sequence length causes the disparate tendency of pooled values in both steps with degradation of performance. To circumvent this problem, in testing steps, utterances are divided by $T$, which is defined in training steps, then FA is applied to each divided segments. For example, if $T$ in training steps is 50 time steps, FA is applied every 50 time steps in test data, e.g. 1-50 time steps, 51-100 time steps, etc. The last left over steps are pooled by themselves.

\subsection{Dual Attention 1 (DA-1)}
\label{ssec_DA1}
To exploit both temporal and frequential information simultaneously, the dual attention 1 (DA-1) is suggested. The process of DA-1 is illustrated in Figure \ref{fig:DA-1}.
DA-1 uses hidden feature map $H$ directly and convolves it by 2-dimensional convolution layers (step \raisebox{.5pt}{\textcircled{\raisebox{-.9pt} {1}}} - \raisebox{.5pt}{\textcircled{\raisebox{-.9pt} {3}}}). Merging method is element-wise summation like in TA and FA. 
\begin{gather}
H_{dual} = f_{dual}(H)
\\H' = H \oplus \sigma(H_{dual})\quad for\quad H_{dual} \in \mathbb{R}^{T \times D}\,,
\end{gather}
where $f_{dual}$ denotes the 2-dimensional convolution in the DA-1 module with kernel size of 7. It consists of 3 layers and the number of filters is 1, 3 and 1, respectively.

\subsection{Dual Attention 2 (DA-2)}
\label{ssec_DA2}

The other way for exploiting both temporal and frequential information is using $H_{temp}$ and $H_{freq}$ at once in summation, called dual attention 2 (DA-2). It takes advantages of both TA and FA modules by combining them. It shows the best performance in 4 kinds of proposed attention modules. The activation function and merging method are same as in TA and FA.
\begin{gather}
H' = H \oplus \sigma(H_{temp} \oplus H_{freq}) \,.
\end{gather}
Computations for obtaining $H_{temp}$ ($h_{temp}$) and $H_{freq}$ ($h_{freq}$), in Eq. (\ref{eq:TA}) and Eq. (\ref{eq:FA}), are executed in parallel.

Every convolution operation in proposed attention modules is followed by batch normalization \cite{Ioffe2015} and ReLU activation function. However, in the very last layer of attention module, batch normalization and activation function are not used because of using sigmoid function before merging. Attention modules are applied after every hidden feature map, even for hidden feature map from last LSTM layer. Also, same attention module is shared across all hidden feature maps from different LSTM layers. It means there is no need to train several attention modules as many as the number of hidden layers in LSTM.

\section{Loss Functions}
Since it is hard to record audio samples in equal or similar ratio of speech to non-speech, imbalance between speech and non-speech sections can be found easily in lots of datasets. To balance the ratio of speech to non-speech for VAD tasks, many researchers manipulate the data by artificially appending silence segments at the beginning and the end of audio samples in training datasets \cite{Kim2016, Jung2018, Graf2015, Drugman2016, Yoo2015, Ghosh2011, Aneeja2015}. To avoid this inconvenience, we utilize the focal loss, revised version of cross entropy loss \cite{Lin2017}.

\subsection{Cross Entropy Loss}
Cross entropy loss is represented as below.
\begin{gather}
l_{CE}(y_{t}) = - \log (y_{t})\\ y_{t}=\left\{\begin{array}{cc}\ \hat{y} & \mbox{if\quad$y = 1$} \\ 1-\hat{y} & \mbox{otherwise}\end{array}\right.\label{eq:yt}\,,
\end{gather}
where $y$ is label and $\hat{y}$ is model's predicted probability for label $y=1$. Thanks to its convexity in optimization, it is widely used in deep learning-based task. In spite of its usefulness, cross entropy loss is hard to naturally handle the inefficient training caused by class imbalance.

\subsection{Focal Loss}
To mitigate the inefficient training in class-imbalanced environment, focal loss is suggested and described as below.
\begin{gather}
l_{FL}(y_{t})=-(1-y_{t})^\gamma \log (y_{t})\label{eq:FL}\,,
\end{gather}
where $\gamma$ is tunable parameter named \emph{focusing} parameter and $y_{t}$ is same with in cross entropy loss, Eq. (\ref{eq:yt}). In focal loss, the modulating factor $(1-y_{t})^\gamma$ is multiplied to cross entropy loss. Modulating factor is increased when the difference between target and predicted value is increased (misclassified case). Otherwise, when the difference is decreased, modulating factor is also decreased (well-classified case). From this mechanism, it strengthens the significance of correcting misclassified examples and alleviates the bias oriented to dominating class.

\begin{table}[t!]
\centering
\caption{Details about LSTM setting of baseline models.}
\vspace*{-0.3cm}
\begin{scriptsize}
\begin{tabular}{@{}ccc@{}}
\toprule
Model             & \# hidden layers & \# hidden units per layer \\
\midrule
\textit{LSTM\_64} & 3                & 64                        \\
\textit{LSTM\_96} & 3                & 96                        \\
\textit{CLDNN\_64\cite{Zazo2016}} & 2                & 64                        \\
\textit{CLDNN\_80\cite{Zazo2016}} & 3                & 80                        \\ \bottomrule
\end{tabular}
\end{scriptsize}
\vspace{-0.5cm}
\label{tab:Baseline}
\end{table}

\section{Experimental Setup}

\subsection{Datasets}
The experiments were conducted on Aurora 4 \cite{Parihar2002} which contains 7,138 and 330 clean utterances for training and testing, respectively. All the clean utterances of training data were corrupted by the public 100 noise types \footnote{web.cse.ohio-state.edu/pnl/corpus/HuNonspeech/HuCorpus.html} at SNR from -10 to 15 dB in 5 dB steps. Noise types and SNRs were selected randomly. This procedure was repeated until training sets reached about 60 hours long. To evaluate the performance in mismatched noisy conditions, we added 5 unseen noises (babble, destroyer-engine, F16 cockpit, factory, and street) with 4 SNRs (-5, 0, 5, and 10 dB) to all of testing data. Because Aurora 4 data show speech dominated class imbalance, 1 second of silence were inserted at back and forth of each utterance in training sets (\textit{1 sec padding}). 

To do experimental work for focal loss, we used training sets without silence padding (\textit{no padding}) and manipulated them for making various imbalance situations. At first, a kind of endpoint detection was executed based on ground-truth (\textit{EPD}). That is to say, the front part before first speech frame and the latter part after last speech frame were deleted. For making opposite condition, we inserted the silence at back and forth of audio samples for 2 seconds and 3 seconds (\textit{2 sec padding} and \textit{3 sec padding}, respectively). The \emph{focusing} parameter $\gamma$ of focal loss in Eq. (\ref{eq:FL}) was set as 0.2, 0.4, 0.6, 0.8, 1.0, 2.0, and 3.0.

\subsection{Setting}
40-dimensional log Mel-filterbanks were used as acoustic features with 25-ms frame length and 10-ms shift length. The ground-truth of noisy speech was extracted by applying Sohn VAD \cite{Sohn1999} to corresponding clean speech. For proving effectiveness of proposed attention module, we used 2 basic LSTM models (\textit{LSTM\_64} and \textit{LSTM\_96}) and 2 CLDNN (Convolutional, Long Short-Term Memory, Deep Neural Networks) models (\textit{CLDNN\_64} and \textit{CLDNN\_80}), the combination of CNN and LSTM, proposed in \cite{Zazo2016}. Model details about LSTM can be seen in Table \ref{tab:Baseline} and remaining details of CLDNNs were same with \cite{Zazo2016}. For finding the best attention module, all of proposed attention modules were integrated to \textit{LSTM\_64}. After finding the best attention module, the rest of baseline models were compared. All models were trained using stochastic gradient descent (SGD) for 20 epochs with an initial learning rate $lr=10^{-1}$ using a batch-size of 128. $lr$ is reduced by a factor of $10^{-1}$ with $10^{-5}$ of lower bound. The LSTM is unrolled for 50 time steps in training to include long-term dependency with truncated backpropagation through time (BPTT).

\begin{table}[t!]
\caption{Averaged AUC(\%) of 5 noises and number of parameters. In this paper, the best results are highlighted in bold and \emph{RI} with parenthesis represents the relative improvement (except for \emph{Table \ref{tab:exp3}}).}
\vspace*{-0.3cm}
\begin{scriptsize}
\begin{tabular}{@{}c||c|c|c|c|c@{}}
\toprule
SNR   & \textit{LSTM\_64} & w/ TA    & w/ FA    & w/ DA-1    & w/ DA-2        \\ \midrule
-5 dB & 87.05             & 88.37 & 89.38 & 88.33 & \textbf{90.06} \\
0 dB  & 94.13             & 94.92 & 94.89 & 94.85 & \textbf{95.42} \\
5 dB  & 97.42             & 97.77 & 97.67 & 97.77 & \textbf{97.90} \\
10 dB & 98.74             & 98.82 & 98.83 & 98.88 & \textbf{98.93} \\ \hline
\begin{tabular}[c]{@{}c@{}}Avg.\\ (RI)\end{tabular} &
  \begin{tabular}[c]{@{}c@{}}94.33\\ (-)\end{tabular} &
  \begin{tabular}[c]{@{}c@{}}94.97\\ (11.29 \%)\end{tabular} &
  \begin{tabular}[c]{@{}c@{}}95.19\\ (15.17 \%)\end{tabular} &
  \begin{tabular}[c]{@{}c@{}}94.96\\ (11.11 \%)\end{tabular} &
  \textbf{\begin{tabular}[c]{@{}c@{}}95.58\\ (22.05 \%)\end{tabular}} \\ 
\hline\hline
\begin{tabular}[c]{@{}c@{}}\# Param.\\ (Increase)\end{tabular} &
  \begin{tabular}[c]{@{}c@{}}95,809\\ (-)\end{tabular} &
  \begin{tabular}[c]{@{}c@{}}96,627\\ (0.85\%)\end{tabular} &
  \begin{tabular}[c]{@{}c@{}}97,327\\ (1.58 \%)\end{tabular} &
  \begin{tabular}[c]{@{}c@{}}96,565\\ (0.79 \%)\end{tabular} &
  \begin{tabular}[c]{@{}c@{}}98,145\\ (2.44 \%)\end{tabular} \\ \bottomrule
\end{tabular}
\end{scriptsize}
\vspace{-0.5cm}
\label{tab:exp1}
\end{table}
\section{Results}

\begin{table*}[t]
\renewcommand\thetable{4}
\centering
\caption{Averaged AUC(\%) of 5 noises in all SNRs for the baseline (\textit{LSTM\_64}) and \emph{DA-2} based model. \emph{CE} and \emph{FL} denote cross entropy and focal loss, respectively. Value in parenthesis after \emph{FL} is the focusing parameter $\gamma$. Results which outperform the \emph{CE}-based result are highlighted in bold. The bottom row represents the ratio of speech (\emph{S}) to non-speech (\emph{NS}) of training data in each situation.}
\vspace*{-0.3cm}
\begin{scriptsize}
\begin{tabular}{@{}c||cc|cc|cc|cc|cc@{}}
\toprule
\multirow{2}{*}{Loss ($\gamma$)} &
  \multicolumn{2}{c|}{\textit{EPD}} &
  \multicolumn{2}{c|}{\textit{no padding}} &
  \multicolumn{2}{c|}{\textit{1 sec padding}} &
  \multicolumn{2}{c|}{\textit{2 sec padding}} &
  \multicolumn{2}{c}{\textit{3 sec padding}} \\ \cline{2-11} 
 &
  Baseline &
  w/ DA-2 &
  Baseline &
  w/ DA-2 &
  Baseline &
  w/ DA-2 &
  Baseline &
  w/ DA-2 &
  Baseline &
  w/ DA-2 \\ \cmidrule(r){1-11}
CE (-) &
  91.70 &
  93.45 &
  92.64 &
  94.81 &
  94.33 &
  95.58 &
  94.47 &
  95.42 &
  94.38 &
  95.33 \\
FL (0.2) &
  \textbf{92.33} &
  \textbf{93.82} &
  \textbf{93.30} &
  94.64 &
  \textbf{94.40} &
  95.43 &
  \textbf{94.64} &
  \textbf{95.49} &
  \textbf{94.53} &
  95.23 \\
FL (0.4) &
  \textbf{92.33} &
  93.33 &
  \textbf{93.18} &
  94.04 &
  93.39 &
  95.39 &
  \textbf{94.58} &
  \textbf{95.50} &
  \textbf{94.61} &
  95.22 \\
FL (0.6) &
  \textbf{92.27} &
  93.34 &
  \textbf{93.04} &
  94.04 &
  \textbf{94.40} &
  95.52 &
  \textbf{94.64} &
  95.41 &
  \textbf{94.41} &
  \textbf{95.49} \\
FL (0.8) &
  \textbf{92.23} &
  93.44 &
  \textbf{92.91} &
  \textbf{95.40} &
  \textbf{94.39} &
  \textbf{95.59} &
  \textbf{94.55} &
  95.41 &
  \textbf{94.47} &
  \textbf{95.46} \\
FL (1.0) &
  \textbf{92.18} &
  92.88 &
  92.55 &
  \textbf{94.88} &
  \textbf{94.39} &
  95.57 &
  \textbf{94.53} &
  \textbf{95.48} &
  \textbf{94.40} &
  \textbf{95.35} \\
FL (2.0) &
  \textbf{91.78} &
  92.62 &
  92.31 &
  94.54 &
  94.29 &
  95.50 &
  94.27 &
  95.21 &
  94.18 &
  95.19 \\
FL (3.0) &
  91.38 &
  92.19 &
  92.19 &
  93.95 &
  94.09 &
  95.52 &
  94.20 &
  94.56 &
  93.99 &
  94.97 \\ \hline\hline
Ratio (S / NS) &
  \multicolumn{2}{c|}{69.96 / 30.04} &
  \multicolumn{2}{c|}{61.24 / 38.76} &
  \multicolumn{2}{c|}{48.59 / 51.41} &
  \multicolumn{2}{c|}{40.23 / 59.77} &
  \multicolumn{2}{c}{34.32 / 65.68} \\ \bottomrule
\end{tabular}
\end{scriptsize}
\label{tab:exp3}
\vspace{-0.35cm}
\end{table*}

\subsection{Comparison of different attention modules}
Table \ref{tab:exp1} represents the results of the baseline (\textit{LSTM\_64}) and baselines integrated with all of proposed attention modules. Evaluation metric is the area under the ROC curve (AUC) \cite{Hanley1982}. The results of 5 noises are averaged along same SNR level and the number of parameters is also compared. 

From this table, we can observe that all of attention-based models outperform the baseline. Also, the increase in number of parameters in all of attention-based models is under 2.5 \%, which is negligible. First of all, the attention even only for single domain, the frequential attention (FA) or temporal attention (TA), can help LSTM model to be optimized in VAD. In -5 dB SNR, FA outperforms the TA. However, both show similar results in other SNR levels. It implies that attention in frequency domain is more effective than in time domain especially in desperately noisy environment, under 0 dB SNR.

\begin{figure}[t!]
\centering
\hspace*{-.55cm}
\centerline{\includegraphics[clip, trim=0.0cm 0.0cm 0.0cm 0.0cm, width=\linewidth]{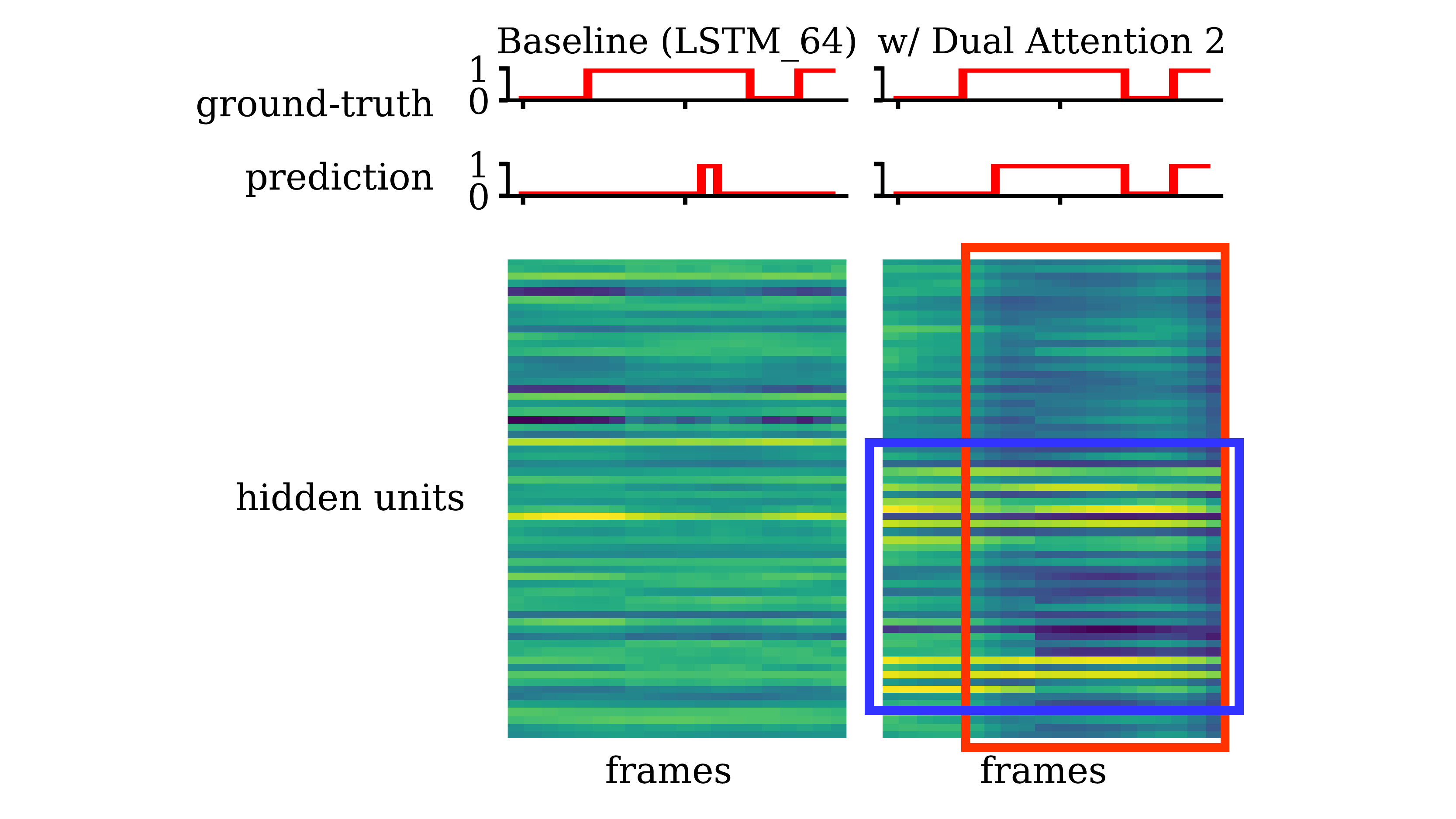}}
\vspace*{-0.3cm}
\caption{The last hidden layer's feature map of 20 frames from 446c0201.wav: baseline and dual attention-based model.}
\label{fig:hidden_feature_map}
\vspace{-0.5cm}
\end{figure}

\begin{table}[b!]
\vspace{-0.5cm}
\renewcommand\thetable{3}
\caption{Averaged AUC(\%) of 5 noises and the number of parameters for other baseline models and dual attention 2 (DA-2).}
\vspace*{-0.3cm}
\centering
\begin{scriptsize}
\begin{tabular}{@{}c|c||c|c@{}}
\toprule
\multicolumn{2}{c||}{Model}                      & Avg. (RI)                 & \# Param. (Increase) \\ \hline
\multirow{3}{*}{Baseline} & \textit{LSTM\_96}  & 94.42 (-)                 & 205,121 (-)        \\
                          & \textit{CLDNN\_64} & 94.53 (-)                 & 129,883 (-)        \\
                          & \textit{CLDNN\_80} & 94.55 (-)                 & 215,927 (-)        \\
\midrule
\multirow{3}{*}{\begin{tabular}[c]{@{}c@{}}Attention\\(w/ DA-2)\end{tabular}} & \textit{LSTM\_96} & \textbf{95.38 (17.20 \%)} & 207,457 (1.14 \%) \\
                          & \textit{CLDNN\_64} & \textbf{95.55 (18.65 \%)} & 132,219 (1.80 \%)   \\
                          & \textit{CLDNN\_80} & \textbf{95.42 (15.96 \%)}  & 218,263 (1.08 \%)   \\ \bottomrule 
\end{tabular}
\label{tab:exp2}
\end{scriptsize}
\end{table}
Although both of the dual attention 1 (DA-1) and the dual attention 2 (DA-2) utilize temporal and frequential information, DA-1 based model shows the lowest results among the attention modules in averaged AUC while DA-2 based model shows the best results in whole SNR levels. It is because DA-1 only focuses on local information by using 2-dimensional convolution on hidden states directly. On the other hand, in DA-2, statistically pooled vectors contain overall information of hidden states. In addition, DA-2 uses both information more precisely by using 1-dimensional convolution separately in each domain. However, DA-1 outperforms baseline as well and uses the least number of parameters among the attention modules. 

For showing effectiveness of DA-2, test waveform sample \emph{446c0201.wav}, corrupted by F16 cockpit with 0 dB SNR, was selected. Afterward, consecutive 20 frames were selected randomly from this sample. The above 2 graphs of Figure \ref{fig:hidden_feature_map} represent ground-truth and predicted results of selected 20 frames (1:speech / 0:non-speech). The left column of below color map shows the hidden feature map of last LSTM layer from baseline. The color map of right column shows the refined hidden feature map of last LSTM layer from DA-2 based model. The darker the color, the greater the activation value. DA-2 based model concentrates on time steps of speech frame and suppress time steps of non-speech frame by TA (indicated by the red rectangular). In addition, unlike the values of each hidden units are similar in baseline, the differences of values are distinct in DA-2 based model (indicated by the blue rectangular). It means FA strengthens the specific hidden units in helpful way to improve the model's ability. In DA-2 based model, the hidden feature map is adaptively refined by TA and FA. As a result, DA-2 based model shows more accurate prediction rather than baseline.

The results of extended experiments to other 3 baseline models (\textit{LSTM\_96, CLDNN\_64, and CLDNN\_80}) are reported in Table \ref{tab:exp2}. We find that DA-2 module improves the performance in all of baselines, even in CLDNN. It means DA-2 module can be flexibly integrated to LSTM and CLDNN-based models with a small increase in number of parameters.

\subsection{Focal loss for various imbalance situations}

Table \ref{tab:exp3} describes the results of experiment about focal loss. \textit{LSTM\_64} was used as baseline and compared with DA-2 based model. The bottom row of table represents the ratio of speech to non-speech in each situation. Ratio of speech is decreased as column of table is moved from left to right. It can be found that focal loss is effective in all of situations, even in balanced situation (\emph{1 sec padding}). In speech dominated situations (\emph{EPD} and \emph{no padding}), focal loss shows more improved results than in opposite situations (\emph{2 sec padding} and \emph{3 sec padding}). When $\gamma=0.2$, result of \emph{no padding} in baseline is 93.30 \%, the 7.59 \% relative improvement above cross entropy loss. Whereas, result of \emph{3 sec padding} in baseline is 94.53 \%, only the 2.67 \% relative improvement above cross entropy loss. Also, the effect of focal loss is less remarkable in DA-2 based model generally.

Table \ref{tab:mean_and_std} represents the mean and standard deviation of cross entropy-based results (the top row in Table \ref{tab:exp3}) from 5 different padding situations. For comparing baseline and DA-2 based model in imbalance situations, mean and standard deviation are obtained along same model. DA-2 based model shows superior result with 21.85 \% of relative improvement on average. It means proposed attention module also takes effect in imbalance situations. Standard deviation is decreased with 30.97 \% in DA-2 based model. It implies that DA-2 based model is more stable than baseline in various imbalance situations.

\begin{table}[t!]
\centering
\caption{Mean and standard deviation of the top row in Table \ref{tab:exp3}.}
\vspace*{-0.3cm}
\begin{scriptsize}
\begin{tabular}{@{}ccc@{}}
\toprule
Model          & Mean (RI)  & Standard deviation (Increase) \\ \midrule
Baseline (\textit{LSTM\_64})  & 93.50 (-) & 1.13 (-)              \\
Attention (w/ DA-2)& 94.92 (21.85 \%) & 0.78 (-30.97 \%)              \\ \bottomrule
\end{tabular}
\end{scriptsize}
\label{tab:mean_and_std}
\vspace{-0.5cm}
\end{table}

\section{Conclusion}
This paper proposed a novel VAD model applying dual attention module which exploits the time and frequency information and infers optimal attention vectors for each domain adaptively. As a result, the proposed attention module improves the performance compared to baseline in unseen noise environment with a slight increase in number of parameters. In addition, the proposed attention module can be flexibly integrated to other LSTM-based baselines for better performance. Additionally, by using focal loss in diverse imbalance situations, performance degradation is alleviated compared to using cross entropy loss.

\section{Acknowledgements}

This work was conducted by Center for Applied Research in Artificial Intelligence (CARAI) grant funded by DAPA and ADD (UD190031RD).

\bibliographystyle{IEEEtran}

\bibliography{mybib}

\begin{thebibliography}{10}
\providecommand{\url}[1]{#1}
\csname url@samestyle\endcsname
\providecommand{\newblock}{\relax}
\providecommand{\bibinfo}[2]{#2}
\providecommand{\BIBentrySTDinterwordspacing}{\spaceskip=0pt\relax}
\providecommand{\BIBentryALTinterwordstretchfactor}{4}
\providecommand{\BIBentryALTinterwordspacing}{\spaceskip=\fontdimen2\font plus
\BIBentryALTinterwordstretchfactor\fontdimen3\font minus
  \fontdimen4\font\relax}
\providecommand{\BIBforeignlanguage}[2]{{%
\expandafter\ifx\csname l@#1\endcsname\relax
\typeout{** WARNING: IEEEtran.bst: No hyphenation pattern has been}%
\typeout{** loaded for the language `#1'. Using the pattern for}%
\typeout{** the default language instead.}%
\else
\language=\csname l@#1\endcsname
\fi
#2}}
\providecommand{\BIBdecl}{\relax}
\BIBdecl

\bibitem{Rabiner1975}
L.~R. Rabiner and M.~R. Sambur, ``{An algorithm for determining the endpoints
  of isolated utterances},'' \emph{The Bell System Technical Journal}, vol.~54,
  no.~2, pp. 297--315, 1975.

\bibitem{Junqua1992}
J.~C. Junqua, B.~Reaves, and B.~Mak, ``{A study of endpoint detection
  algorithms in adverse conditions: Incidence on a DTW and HMM recognize},'' in
  \emph{Proc. of EUROSPEECH}, 1991, pp. 1371--1374.

\bibitem{Haigh1993}
J.~A. Haigh and J.~S. Mason, ``{Robust voice activity detection using cepstral
  features},'' in \emph{Proc. of TENCON '93. IEEE Region 10 International
  Conference on Computers, Communications and Automation}, 1993, pp. 321--324.

\bibitem{Shen1998}
J.~Shen, J.~Hung, and L.~Lee, ``{Robust entropy-based endpoint detection for
  speech recognition in noisy environments},'' in \emph{Proc. of International
  Conference on Spoken Language Processing (ICSLP)}, 1998, pp. 232--235.

\bibitem{Sohn1999}
J.~Sohn, N.~S. Kim, and W.~Sung, ``{A statistical model-based voice activity
  detection},'' \emph{IEEE Signal Processing Letters}, vol.~6, no.~1, pp. 1--3,
  1999.

\bibitem{Chang2006}
J.~H. Chang, N.~S. Kim, and S.~K. Mitra, ``{Voice activity detection based on
  multiple statistical models},'' \emph{IEEE Transactions on Signal
  Processing}, vol.~54, no.~6, pp. 1965--1976, 2006.

\bibitem{Zhang2014}
X.~L. Zhang and D.~L. Wang, ``Boosted deep neural networks and
  multi-resolutiton cochleagram features for voice activity detection,'' in
  \emph{Proc. of Interspeech}, 2014, pp. 1534--1538.

\bibitem{Jung2017}
Y.~Jung, Y.~Kim, H.~Lim, and H.~Kim, ``Linear-scale filterbank for deep neural
  network-based voice activity detection,'' in \emph{Proc. of Conference of the
  Oriental Chapter of International Committee for Coordination and
  Standardization of Speech Databases and Assessment Technique (O-COCOSDA)},
  2017, pp. 43--47.

\bibitem{Jung2018}
Y.~Jung, Y.~Kim, Y.~Choi, and H.~Kim, ``Joint learning using denoising
  variational autoencoders for voice activity detection,'' in \emph{Proc. of
  Interspeech}, 2018, pp. 1210--1214.

\bibitem{Thomas2014}
S.~Thomas, S.~Ganapathy, G.~Saon, and H.~Soltau, ``{Analyzing convolutional
  neural networks for speech activity detection in mismatched acoustic
  conditions},'' in \emph{Proc. of the IEEE International Conference on
  Acoustics, Speech and Signal Processing (ICASSP)}, 2014, pp. 2519--2523.

\bibitem{Sehgal2018}
A.~Sehgal and N.~Kehtarnavaz, ``{A convolutional neural network smartphone app
  for real-time voice activity detection},'' \emph{IEEE Access}, vol.~6, pp.
  9017--9026, 2018.

\bibitem{Lin2019}
R.~Lin, C.~Costello, C.~Jankowski, and V.~Mruthyunjaya, ``{Optimizing voice
  activity detection for noisy conditions},'' in \emph{Proc. of Interspeech},
  2019, pp. 2030--2034.

\bibitem{Eyben2013}
F.~Eyben, F.~Weninger, S.~Squartini, and B.~Schuller, ``{Real-life voice
  activity detection with LSTM recurrent neural networks and an application to
  hollywood movies},'' in \emph{Proc. of the IEEE International Conference on
  Acoustics, Speech and Signal Processing (ICASSP)}, 2013, pp. 483--487.

\bibitem{Kim2016}
J.~Kim, J.~Kim, S.~Lee, J.~Park, and M.~Hahn, ``{Vowel based voice activity
  detection with LSTM recurrent neural network},'' in \emph{Proc. of the
  International Conference on Signal Processing Systems (ICSPS)}, 2016, pp.
  134--137.

\bibitem{Sertsi2017}
P.~Sertsi, S.~Boonkla, V.~Chunwijitra, N.~Kurpukdee, and C.~Wutiwiwatchai,
  ``{Robust voice activity detection based on LSTM recurrent neural networks
  and modulation spectrum},'' in \emph{Proc. of the Asia Pacific Signal and
  Information Processing Association (APSIPA)}, 2017, pp. 342--346.

\bibitem{Zazo2016}
R.~Zazo, T.~N. Sainath, G.~Simko, and C.~Parada, ``{Feature learning with
  raw-waveform CLDNNs for voice activity detection},'' in \emph{Proc. of
  Interspeech}, 2016, pp. 3668--3672.

\bibitem{Vafeiadis2019}
A.~Vafeiadis, E.~Fanioudakis, I.~Potamitis, K.~Votis, D.~Giakoumis,
  D.~Tzovaras, L.~Chen, and R.~Hamzaoui, ``{Two-dimensional convolutional
  recurrent neural networks for speech activity detection},'' in \emph{Proc. of
  Interspeech}, 2019, pp. 2045--2049.

\bibitem{Hu2017}
J.~Hu, L.~Shen, and G.~Sun, ``{Squeeze-and-excitation networks},'' in
  \emph{Proc. of the IEEE Conference on Computer Vision and Pattern Recognition
  (CVPR)}, 2018, pp. 7132--7141.

\bibitem{Woo2018}
S.~Woo, J.~Park, J.~Lee, and I.~Kwon, ``{CBAM: Convolutional block attention
  module},'' in \emph{Proc. of the European Conference on Computer Vision
  (ECCV)}, 2018, pp. 3--19.

\bibitem{Yadav2020}
S.~Yadav and A.~Rai, ``{Frequency and temporal convolutional attention for
  text-independent speaker recognition},'' in \emph{Proc. of the IEEE
  International Conference on Acoustics, Speech and Signal Processing
  (ICASSP)}, 2020, pp. 6794--6798.

\bibitem{Yanpei2020}
Y.~{Shi}, Q.~{Huang}, and T.~{Hain}, ``{Robust speaker recognition using speech
  enhancement and attention model},'' \emph{arXiv e-prints}, p.
  arXiv:2001.05031, 2020.

\bibitem{Japkowicz2002}
N.~Japkowicz and S.~Stephen, ``{The class imbalance problem: A systematic
  study},'' \emph{Intelligent Data Analysis}, vol.~6, no.~5, pp. 429--449,
  2002.

\bibitem{He2009}
H.~He and E.~A. Garcia, ``{Learning from imbalanced data},'' \emph{IEEE
  Transactions on knowledge and data engineering}, vol.~21, no.~9, pp.
  1263--1284, 2009.

\bibitem{Lin2017}
T.~Y. Lin, P.~Goyal, R.~Girshick, K.~He, and P.~Doll\'{a}r, ``{Focal loss for
  dense object detection},'' in \emph{Proc of the IEEE International Conference
  on Computer Vision (ICCV)}, 2017, pp. 2980--2988.

\bibitem{Tong2016}
S.~Tong, H.~Gu, and K.~Yu, ``{A comparative study of robustness of deep
  learning approaches for VAD},'' in \emph{Proc. of the IEEE International
  Conference on Acoustics, Speech and Signal Processing (ICASSP)}, 2016, pp.
  5695--5699.

\bibitem{Wang2019}
M.~Wang, Q.~Huang, J.~Zhang, Z.~Li, H.~Pu, J.~Lei, and L.~Wang, ``{Deep
  learning approaches for voice activity detection},'' in \emph{Proc. of the
  International Conference on Cyber Security Intelligence and Analytics
  (CSIA)}, 2019, pp. 816--826.

\bibitem{Sepp1997}
S.~Hochreiter and J.~Schmidhuber, ``{Long short-term memory},'' \emph{Neural
  Computation}, vol.~9, no.~8, pp. 1735--1780, 1997.

\bibitem{Ioffe2015}
S.~Ioffe and C.~Szegedy, ``Batch normalization: Accelerating deep network
  training by reducing internal covariate shift,'' in \emph{Proc. of
  International Conference on Machine Learning (ICML)}, 2015, pp. 448--456.

\bibitem{Graf2015}
S.~Graf, T.~Herbig, M.~Buck, and G.~Schmidt, ``{Features for voice activity
  detection: a comparative analysis},'' \emph{EURASIP Journal on Advances in
  Signal Processing}, vol. 2015, no.~91, pp. 1--15, 2015.

\bibitem{Drugman2016}
T.~Drugman, Y.~Stylianou, Y.~Kida, and M.~Akamine, ``{Voice activity detection:
  merging source and filter-based information},'' \emph{IEEE Signal Processing
  Letters}, vol.~23, no.~2, pp. 252--256, 2016.

\bibitem{Yoo2015}
I.~C. Yoo, H.~Lim, and D.~Yook, ``{Formant-based robust voice activity
  detection},'' \emph{IEEE/ACM Transactions on Audio, Speech, and Language
  Processing}, vol.~23, no.~12, pp. 2238--2245, 2015.

\bibitem{Ghosh2011}
P.~K. Ghosh, A.~Tsiartas, and S.~Narayanan, ``{Robust voice activity detection
  using long-term signal variability},'' \emph{IEEE Transactions on Audio,
  Speech, and Language Processing}, vol.~19, no.~3, pp. 600--613, 2011.

\bibitem{Aneeja2015}
G.~Aneeja and B.~Yegnanarayana, ``{Single frequency filtering approach for
  discriminating speech and nonspeech},'' \emph{IEEE/ACM Transactions on Audio,
  Speech, and Language Processing}, vol.~23, no.~4, pp. 705--717, 2015.

\bibitem{Parihar2002}
D.~Pearce and J.~Picone, ``{Aurora working group: DSR front end LVCSR
  evaluation AU/384/02},'' \emph{Inst.for Signal and Information Process.,
  Mississippi State University, Tech. Rep}, vol.~40, p.~94, 2002.

\bibitem{Hanley1982}
J.~A. Hanley and B.~J. McNeil, ``{The Meaning and Use of the Area under a
  Receiver Operating Characteristic (ROC) Curve},'' \emph{Radiology}, vol. 143,
  pp. 29--36, 1982.

\end{thebibliography}

\end{document}